# Organic Solar Cells; Fabrication Technique, Operating Principle, Characterization and Improvement

Fahmi F. Muhammadsharif

Department of Physics, Faculty of Science and Health, Koya University, Koya, Kurdistan Region-F.R., Iraq.
E-mail: fahmi.fariq@koyauniversity.org
Tel.: + 964-7501168841

**Abstract**

Organic solar cells (OSCs) have received a special attention over the past years due to their solution processability, low cost, flexibility and capability of role-to-role production. The power conversion efficiency of these devices has been significantly increased over the past decades from 1% in 1986 to 5% in 2005 and to up to 13% in 2017 thanks to the molecular optimization and the use of non-fullerene acceptors in their active materials. Despite such interesting efficiency, their applications remain limited so far because of instability and short life time of their active layers. It is expected that these obstacles will be surmounted in a foreseeable future upon rigorous research studies performed in the field. This paper is devoted to reviewing the operating principle, characterization parameters and the most important approaches that are considered aiming at improving the overall performance of these devices.

## 1. Introduction

Energy has become a powerful engine of economic, technological, and social development to every country. The rate of energy consumption in the world increases year after year, in which the developed countries can be considered to have the main contribution of this consumption [1]. The continuous demand of energy and the limiting supply of its today's main sources (petroleum, natural gas, and coal) with their detrimental long-term effects on the environment, necessitate a rapid development into the renewable-energy sources. In 2001, the European Union officially recognized the need to promote Renewable Energy Sources (RES) as a priority measure for environmental protection and sustainable development [2]. Among the energies use, electricity is the most versatile form. Access to and consumption of electricity is closely correlated with quality of life. Figure 1.1 shows the Human Development Index (HDI) for over 60 countries, which includes over 90% of the Earth's population, versus the annual per capita electricity use. To improve the quality of life in many countries, as measured by their HDI, increasing their electricity consumption will require by factors of 10 or more, from a few hundred to a few thousand kilowatt-hours (kWh) per year [3]. As such, it is cleared that the usage and neediness for electricity will be on a continuous growth with preceding of time. Therefore, the only sustainable source that can supply this additional electric power to the world with its capability of protecting our environment from being polluted is thought to be the solar energy [4-6]. Earth is receiving solar energy from the sun in one hour with an amount larger than that the world is using it during a whole year [7].

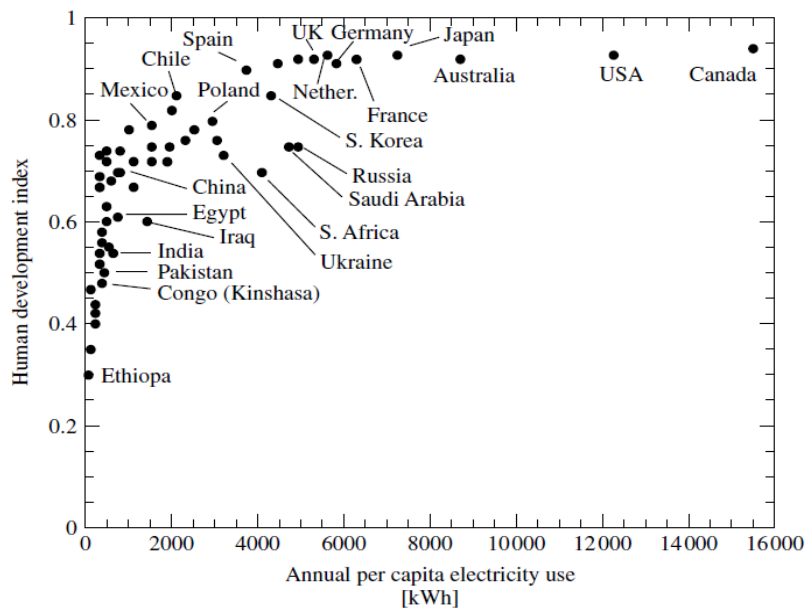

Figure 1.1: Human development index (HDI) versus per capita kWh electricity use.

Organic solar cells are basically made from those materials in which carbon atoms are participating in their chemical structures. OCSs are recognized by their potential as low-cost, light weight, and flexible devices that are capable of converting solar energy directly into electricity. This generation of solar cells is different from traditional inorganic p-n junctions, in which free charge carriers are generated directly upon photon absorption. In OSCs electron-hole pairs are inherently generated in the form of exciton (loosely bound electron-hole) before they dissociate into free charge carriers of electrons and holes. This separation of excitons into free electrons and holes are occurred only at the donor-acceptor boundaries due to the differences in the internal electric fields between the photovoltaic active layers. Almost all OSCs have a planar-layered structure, where the organic light-absorbing layer is sandwiched between two different electrodes. The first electrode must be semi-transparent (usually with > 90% transmittance) for the purpose of light absorption by the absorbing layer. Indium-tin-oxide (ITO) is normally utilized as first electrode, but a very thin metal layer can also be used. The most commonly used second electrode is one of these metals; aluminum, calcium, magnesium, or gold [8]. Figure 2.1 shows a prototype structure of OSC device. These devices promise to open up new markets for solar energy, potentially powering everything from watches and calculators to laptop computers. They are the focus of the world wide intense research efforts owing to their potential applications in the low cost, minimal weight, and flexible electric power sources [9, 10].

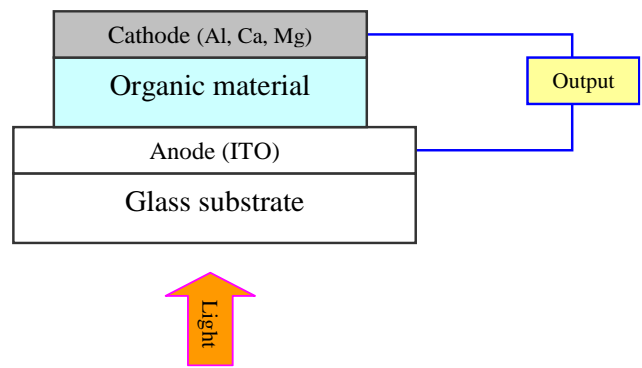

Figure 2.1: Structure of organic based solar cell devices.

## 2. Historical Background

The discovery of photovoltaic (PV) effect in 1839 is referred to Becquerel (Alexandre-Edmond Becquerel, French Physicist), who discovered a photocurrent when platinum electrodes, covered with silver bromide or silver chloride, was illuminated in aqueous solution. Thereafter, Smith and Adams made the first reports on photoconductivity of selenium in 1873 and 1876, respectively. The first organic compound in which its photoconductivity observed was anthracene, investigated by Pochettino in 1906 and Volmer in 1913 [8]. Tang et al. (1987), quoted by [7, 11], at Kodak research laboratories, suggested the first organic solar cell based on the two-layer donor-acceptor concept in 1987. Tang reported a breakthrough in organic photovoltaic performance, achieving power efficiency of nearly 1% under a simulated solar illumination. This solar cell was fabricated by evaporating a 25 nm thin layer of copper phthalocyanine (CuPc) onto an indium tin oxide (ITO)-coated glass substrate, followed by a 45 nm thin layer of 3,4,9,10-perylene tetracarboxylic-bisbenzimidazole (PTCBI) and finally a silver cathode was evaporated on top of the structure. The next convincing step was the application of a dispersed bulk heterojunction of poly(2-methoxy-5(2'-ethyl) hexoxy-phenylenevinylene) (MEH-PPV) polymer and phenyl-$C_{60}$-butyric acid methyl ester ($C_{60}$) and later soluble derivatives of $C_{60}$ (methanofullerene, $C_{61}$), which increased the power conversion efficiency to 2.5% [12]. In 2005, devices with active layers of mixtures of poly-3-hexylthiophene (P3HT) polymer and phenyl-$C_{61}$-butyric acid methyl ester (PCBM) which allowed efficiencies of around 5% has been reached thanks to the nanoscale morphologically control and post production annealing process [13]. Peet *et al*. [14] with the improved light harvesting in the near infrared region by incorporating a low band gap polymer, such as poly[2,6-(4,4-bis-(2-ethylhexyl)-4H-cyclopenta[2,1-b;3,4-b']-dithiophen)-alt-4,7-(2,1,3-benzothiadiazole)] (PCPDTBT), they managed to achieve a higher power conversion efficiency of 5.5%. More recently, a power conversion efficiency (PCE) of about 6% was reported for devices based on fluorinated thieno[3,4-*b*] thiophene and benzodithiophene units PTB4/PC$_{61}$BM films prepared from mixed solvents [15]. In 2011, Chu et al. [16] achieved on power conversion efficiencies of 7.3% by using thieno[3,4-c]pyrrole-4,6-dione and Dithieno[3,2-*b*:20,30-*d*]silole copolymer as active layers. All these achievements proved that polymeric based OSCs have a bright future. However, these values are still far away for daily applications. Table 2.1 shows some important milestones in the development of OSCs [12-19].

Table 1: Some notable events in the history of organic solar cells

| Year | Event |
|---|---|
| 1839 | Becquerel observed the photoelectrochemical process. |
| 1906 | Pochettino studied the photoconductivity of anthracene. |
| 1958 | Kearns and Calvin worked with magnesium phthalocyanines (MgPh), measuring a photovoltage of 200 mV. |
| 1964 | Delacote observed a rectifying effect when magnesium phthalocyanines (CuPh) was placed between two different metalelectrodes. |
| 1986 | Tang published the first heterojunction PV device. |
| 1991 | Hiramoto made the first dye/dye bulk heterojunction PV by co-sublimation. |
| 1993 | Sariciftci made the first polymer/$C_{60}$ heterojunction device. |
| 1994 | Yu made the first bulk polymer/$C_{60}$ heterojunction PV. |

| 1995 | Yu / Hall made the first bulk polymer/polymer heterojunction PV. |
| --- | --- |
| 2000 | Peters / van Hal used oligomer-$C_{60}$ dyads/triads as the active material in PV cells. |
| 2001 | Schmidt-Mende made a self-organised liquid crystalline solar cell of hexabenzocoronene and perylene and Ramos used double-cable polymers in PV cells. |
| 2001 | Shaheen et al. obtained 2.5% conversion efficiency of organic photovoltaic devices based on a conjugated polymer/methanofullerene blends. |
| 2005 | Li et al. reported 4.4% efficient P3HT/$PC_{61}BM$ based OSC by controlling the active layer growth rate. |
| 2005 | Ma et al. made devices with active layers of P3HT/$PC_{61}BM$ with efficiencies of up to around 5%. |
| 2007 | Peet et al. used PCPDTBT/$PC_{71}BM$ to achieve power conversion efficiency of 5.5%. |
| 2007 | Kim et al. received efficiencies of about 6% upon controlling the nanoscale morphology of P3HT/$PC_{61}BM$ active layer. |
| 2009 | Liang et al. made devices based on fluorinated PTB4/$PC_{61}BM$ films fabricated from mixed solvents – efficiency over 6%. |
| 2011 | Chu et al. used Thieno[3,4-c]pyrrole-4,6-dione and Dithieno[3,2-*b*:20,30-*d*]silole Copolymer to obtain power conversion efficiency of about 7.3% |
| 2017 | Zhao et al. used Molecular optimization approach to get 13% |

## 3. Fabrication Techniques

The high dollar-to-performance ratio of inorganic based solar cells is due to the expensive materials such as silicon or gallium arsenide (GaAs), along with high-cost processes which require high temperatures (400-1400 ºC) and high vacuum conditions. Therefore, new environmental friendly photovoltaic devices are needed to drive costs down to the desired levels and to facilitate their production in the ambient temperature condition. Based on the semiconducting properties of organic molecules, the active materials used for fabricating OSCs are soluble in most of common organic solvents. This makes OCSs possess the potential to be fabricated by spinning and common printing techniques [12]. Besides being easily up scalable on rigid as well as on flexible substrates, they open the route of roll-to-roll production of low cost renewable energy sources [20]. Various coating and printing technologies have been proven their compatibility with semiconducting polymer processing. Figure 2.2 shows some standard techniques used in coating and printing processes of the organic active layers in OSCs.

Nevertheless, to achieve a product viable on the market and competitive with the other available technologies, OCSs have to fulfill the standard requirements: cost, efficiency, and lifetime. This latest generation of OSCs potentially offers a convincing solution to the problem of a high cost which commonly encountered for the previous generation PV technologies. However, low power conversion efficiency, poor operational stability; materials cost and environmental impact are still become the main constrains that prevent this OSCs to be fully commercialized in the market [12, 21, 22].

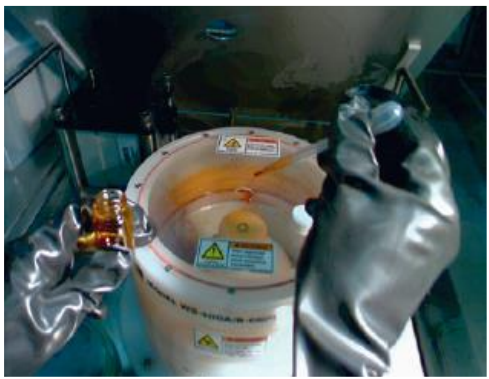

(a)

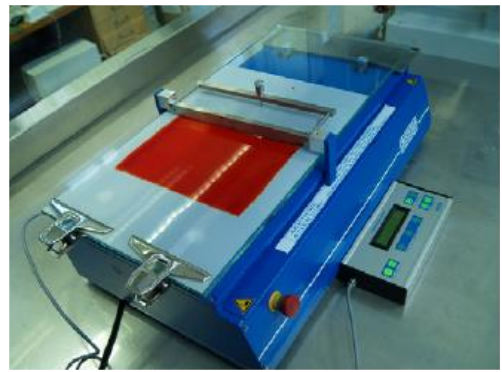

(b)

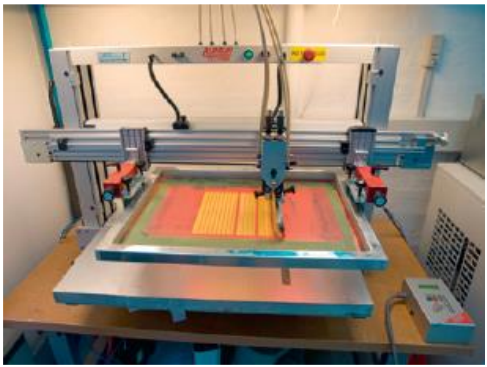

(c)

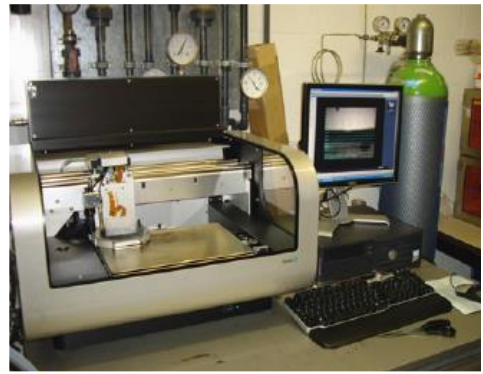

(d)

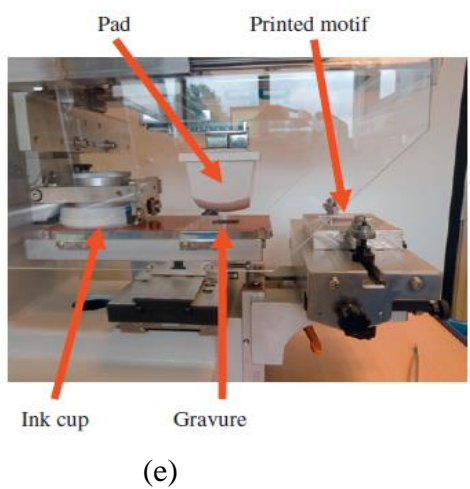 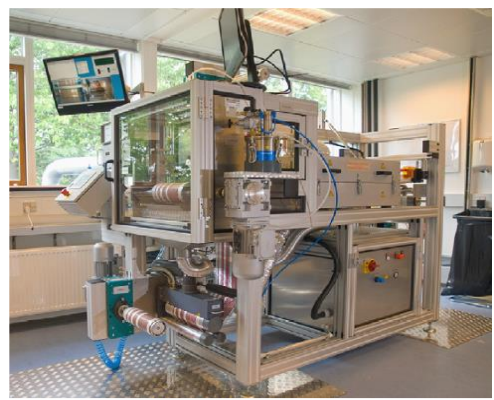

(e)                         (f)

Figure 2.2: Photographic picture of some techniques used for coating and printing active layers of OSC; (a) spin coating, (b) doctor balding, (c) screen printing, (d) ink-jet printing, (e) pad printing and (f) role-to-role technique.

## 4. Physics and Characterization of OSC Devices

Generally, OSC devices can be modeled with an equivalent electrical circuit in the manner that shown in Figure 2.3. The circuit is comprised of the following items:

- A current source ($I_{sc}$): represents the highest photocurrent generated within the cell. This current, flows in opposite direction compared to the forward one of the diode and depends on the voltage across the device. In practical consideration, the value of $I_{sc}$ increases with both of light intensity and temperature [23] and decreases with increasing the thickness layer of the device [24].
- A voltage source ($V$): is facilitated by the voltage drop across the diode, depending on the current passing through it. In the case where the circuit is shorted ($I_{sc}$), its value is zero since no current passes through the diode. Once the circuit is opened, all the current passes through the diode. Hence, the diode potential barrier ($V_f$) limits the voltage to open circuit voltage ($V_{oc}$).
- A series resistance ($R_s$): gathers the ohmic contributions of the electrodes as well as the contact resistance between the organic semiconductor and the metal. Besides, $R_s$ reflects the capability of the organic active layers to transport charge carriers, which has to be minimized as small as possible. The value of $R_s$ is found to be relatively stays unchanged with increasing light intensity [23].
- A shunt resistance [25]: illustrates the potential leakage current through the device. It is the overall quality of the thin film. Unlike the $R_s$, it has to be maximized to reach high efficiency for the device. It has been observed that $R_{sh}$ enlarges with decreasing thickness and reduces drastically with increasing light intensity [23, 26, 27].

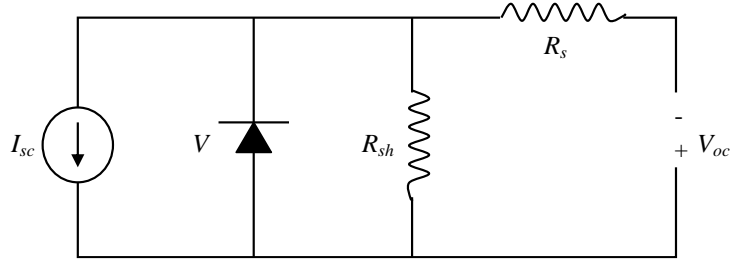

Figure 2.3: An equivalent circuit which models OSC devices.

The current density, $J$ (current per the device active area, $I/A$) of illuminated solar cell is described by the following equation [23]:

$$J = J_o \left[ \exp\left( \frac{q(V - JR_s)}{nk_B T} \right) - 1 \right] - J_L \tag{2.1}$$

where, $J_L$ is the photogenerated current density, $J = J_L \approx J_{sc}$ at $V = 0$ Volt and $J_o$ is referred to as saturation current density and equal to:

$$J_o = N_V N_C \mu k_B T \exp\left( -\frac{E_g}{k_B T} \right) \frac{1}{LN_A} \tag{2.2}$$

where $N_V$, $N_C$ are the effective densities of states in the valence and conduction bands, respectively, $\mu$ is mobility, $E_g$ the band gap, $L$ the mean free path, and $N_A$ the acceptor density.

In order to explain the physics of the devices working principle (light in- current out) and characterization of OSCs, the discussion will be divided into four major sections; first, photo-absorption and exciton formation, second, exciton diffusion and dissociation, third, charge transport and collection at the electrodes, and fourth, devices parameters, which are key factors to get insight into the photovoltaic performance of OSCs.

**4.1 Photo-absorption and Exciton Generation**

In crystalline inorganic semiconductors with a three dimensions (3D) crystal lattice, the individual lower unoccupied molecular orbitals (LUMOs) and higher occupied molecular orbitals (HOMOs) form a conduction band (CB) and valence band (VB) throughout the material, respectively. This is fundamentally different from most organic semiconductors where the intermolecular forces are too weak to form 3D crystal lattices. Consequently, the molecular LUMOs and HOMOs do not contribute strongly enough to form a CB and VB. In conjugated polymers, excitons (bounded electron-hole) are considered to be localized on the specific chain segments. However, there are cases where excitons seem to be delocalized. In these cases, the excitons are referred to as polarons [8, 28].

It is known that the more efficient OSC is one which its thin active layers (the light absorber and charges generator) consists of at least two materials (bi-layer), one is responsible for electron transporting called acceptor (A) and the other for hole transporting called donor (D). As shown in Figure 2.4, when the active layer absorbs light with sufficient photon energy, an electron is excited from the HOMO to the LUMO forming an exciton. Exciton is an electron-hole pair with a zero net charge that is bounded by a weakly coulomb interaction force. In organic PV devices, this process must be followed by exciton dissociation, i.e., the electron-hole pairs must be separated in such as way that electrons should reach the cathode, whilst holes should reach the anode. As a consequence of relatively large exciton binding energies of approximately 200-500 meV [29, 30] and small thermal energy at room temperature (~25 meV) such excitons do not dissociate into free charge carriers quantitatively with thermal excitations. However, the mobilities of organic semiconductors several orders of magnitude less than those of inorganic semiconductors, e.g. electron mobilities for organic materials hovering around 0.1 $cm^2.V^{-1}.s^{-1}$ [29], but strong optical absorption coefficients of organic semiconductors and polymers (~$10^5$ $cm^{-1}$) [31] allow for using <100 nm thin devices, which somehow circumvents the problem of low mobilities.

Therefore, the purposely designed organic materials and tuning their optical band gap for organic solar cells application are crucial to achieve broad absorption spectrum, thereby harvesting sufficient amount of solar energy. Due to these, the low band gap organic materials [14] are usually of prerequisite choices to be utilized as active layer in OSCs. Moreover, the HOMO and LUMO energy levels alignment between the donor and acceptor components is of great importance that would be seen within the next contents of the thesis.

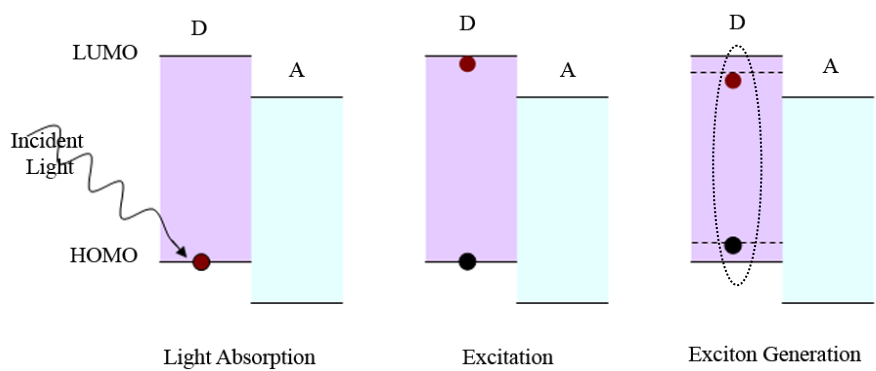

Figure 2.4: Photo-absorption and exciton generation processes in OSC devices.

**4.2 Exciton Diffusion and Dissociation**

Once excitons have been created by the absorption of photons, they can diffuse over a length of approximately 5–15 nm [32, 33] towards the donor-acceptor (D-A) proximity and then dissociate there. It is apparent that exciton dissociation takes place effectively at the interface of the D-A heterojunction, thus the exciton should be formed within the diffusion length of the interface. Figure 2.5 illustrates schematically the charge separation processes at the interface between donor and acceptor in active layer of OSCs. The electron can jump from the LUMO of the donor to the LUMO of the acceptor if the potential difference between the ionization potential (*IP*) of the donor and the electron affinity (*EA*) of the acceptor is larger than the exciton binding energy. However, this process can lead to free charges only if the hole remains on the donor due to its higher HOMO level. In contrast, if the HOMO of the acceptor is higher, the exciton transfers itself completely to the acceptor (i.e., the lower energy gap, $E_g$ material) accompanied by energy loss. Under the right condition, excitons dissociate into free carriers in the time scale of femto-second (~50 fs) [19], while the remaining excitons recombine either radiatively or non radiatively [32, 34]. Recombination is a limiting factor for almost any kind of PV devices, which is the main determinant of $V_{oc}$ and in some cases it may also affect $I_{sc}$ [35].

Thicker film layers increase light absorption but in this case only a small fraction of the excitons will reach the interface and dissociate. This problem can be overcome by blending donor and acceptor, in a concept called dispersed (or bulk) heterojunction that will be discussed later in Section 2.3.1. There are experimental indications for the formation of an interfacial dipole between the donor and acceptor phases due to spontaneous charge transfer across the interfaces [10]. This can stabilize the charge-separated state by a repulsive interaction between the interface and the free charges. Such condition activates the exciton dissociation, hence leading to enhanced charge transfer. Exciton dissociation can also occur at charge traps or impurities, but such films are likely to have poor charge transport [36].

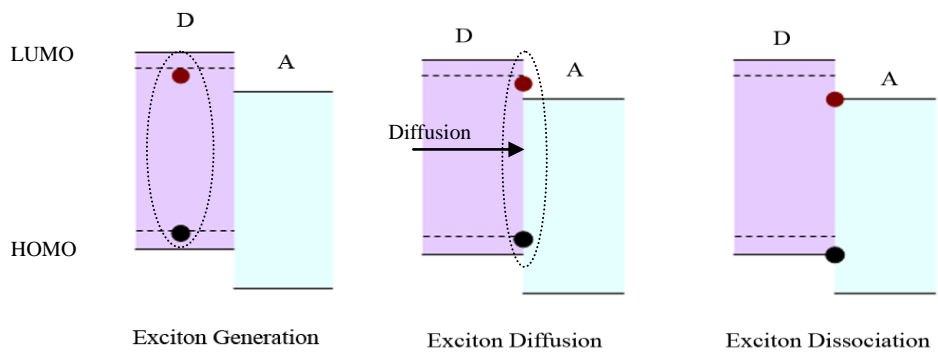

Figure 2.5: Schematic diagram of exciton diffusion and dissociation processes in OSC devices (Black circle represents hole, while red circle represents electron).

**4.3 Charge Transport and Collection**

In the final step of photocurrent generation process and after dissociation of excitons, the free charge carriers undergo the following physical phenomena; either they recombine geminately or transport by diffusion and drift forces through the device, where they recombine with other oppositely charged carriers [30], or they are transported to the appropriate electrodes and out to the external circuit. This final phenomenon leads to the photocurrent generation. Hence, in order to get an efficient device, it needs to be optimized as much as possible. As shown in Figure 2.6, the free charges are swept to the electrodes by a built-in electric field due to the difference in the Fermi energy state relative to the band edges of the p-type (donor) and n-type (acceptor) layers. The built-in electric field arises from the difference in work functions of the cathode (Al) and anode (ITO) electrodes, is insufficient to ionize the Fermi energy state. Alternatively, the photocurrent is generated by a charge transfer reaction between the donor and acceptor molecules due to differences in the electron affinity or ionization potential, or both. However, these considerations which are inherently affecting the open circuit voltage ($V_{oc}$) and performance of the devices are still matters of debate [37, 38]. Since charge transport proceeds by hopping between delocalized states, rather than transport within a band, the charge carrier mobility in organic semiconductors are generally low compared to inorganic semiconductors. In addition, complete charge separation is more difficult in organic semiconductors due to the low dielectric constant [8], which acts upon the medium of the material to polarize less efficiently compared to that of inorganic one. Since charge transfer takes place at the organic heterojunction, absorption must take place at the interface or within the exciton diffusion length in the respective materials. In principle, the exciton diffusion length can be calculated by analyzing the spectral response of a device. Photoluminescence (PL) quenching is a conceptually simple experiment that can be used to estimate diffusion lengths [39, 40].

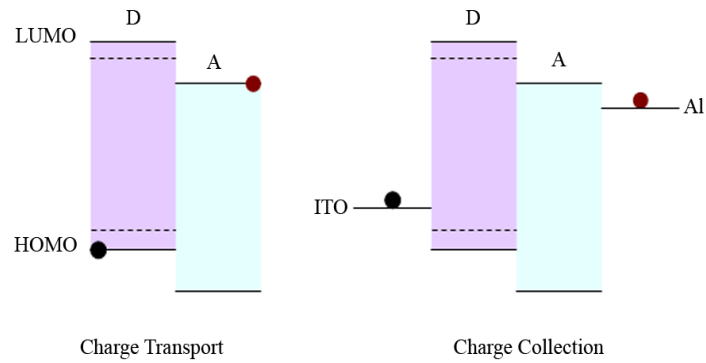

Figure 2.6: Charge transport and collection process in OSC devices.

## 4.4 Characterization Parameters

Figure 2.7 shows a schematic diagram of the current density–voltage curve of a solar cell under illumination. OSC devices are generally characterized by the short-circuit current density ($J_{sc}$), the open-circuit voltage ($V_{oc}$), the fill factor ($FF$) and power conversion efficiency ($\eta$). These four quantities define the performance of a solar cell, and thus they are its key characteristic parameters.

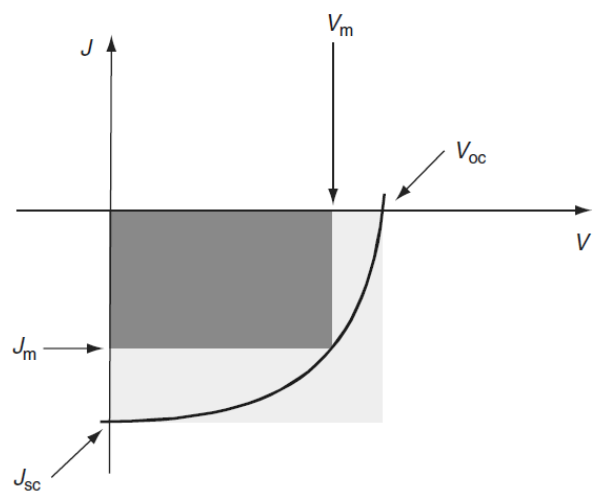

Figure 2.7: *J-V* characteristic of OSC devices.

For the purpose of comparison in the devices performance, all the cells must be considered under a standard illumination condition. The standard test condition (STC) for solar cells, regardless of their design and active material, is the Air Mass 1.5 spectrum (AM 1.5G, represents sunlight with the Sun at an oblique angle of 48.2° above the earth atmosphere), with an incident power density of 1000 W.m$^{-2}$ (100 mW.cm$^{-2}$). This is also defined as the standard 1 sun value, at an ambient temperature of 25 °C. This condition was defined by the American Society for Testing and Materials [41, 42] (see Figure 2.8).

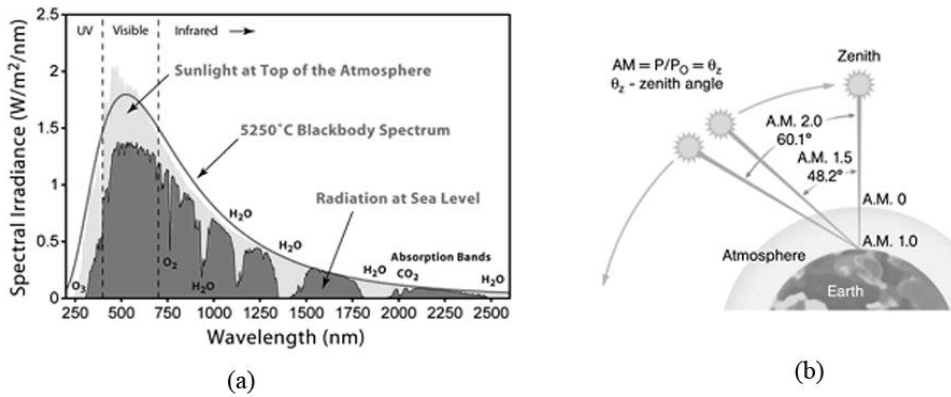

Figure 2.8: shows (a) solar irradiance spectrum above atmosphere and at surface, and (b) air masses at different sun zenith angle.

When the cell is placed in an open circuit and illuminated condition, the electrons and holes separate starting to flow towards the low and high work function electrodes, respectively. At some point, the charge will build up and reach a maximum value equal to the $V_{oc}$, which is limited by the difference in HOMO of the donor and LUMO of the acceptor. The maximum current that can run through the cell is defined by the short-circuit current ($I_{sc}$). This quantity is determined by connecting the two electrodes, whereby the potential across the cell is set to zero, and then illuminating the cell while the current flow is measured simultaneously. $I_{sc}$ yields information about the charge separation and transport efficiency in the cell. The square $I_{max} \times V_{max}$ defines the maximum work (maximum power) that the cell is able to produce. The fill-factor (FF) is given by $I_{max}.V_{max}/V_{oc}.I_{sc}$, and is typically around 0.4–0.6 [8]. The fill factor is the ratio of the dark shaded to the light shaded area under the curve illustrated in Figure 2.7.

$$FF = \frac{I_{max}V_{max}}{I_{sc}V_{oc}} = \frac{P_{max}}{I_{sc}V_{oc}} \qquad (2.3)$$

The power conversion efficiency ($\eta$) of the device is defined as the ratio between the maximum electrical power generated [43] and the incident optical power ($P_{in}$) of photons:

$$\eta = \frac{P_{max}}{P_{in}} = \frac{FF \times I_{sc} V_{oc}}{P_{in}} \quad (2.4)$$

Another important parameter of PV cells is external quantum efficiency (*EQE*), which is defined as the number of generated electrons per incident photon, without correction for reflection loses. Under monochromatic light illumination at a wavelength $\lambda$ (nm), the *EQE* is defined by equation [33]:

$$EQE = \frac{J_{sc} \times hc}{P_o \lambda e} = \frac{1241 J_{sc}}{P_o \lambda} \quad (2.5)$$

and,

$$J_{sc} = \frac{e}{hc} \int_{\lambda_{min}}^{\lambda_{max}} EQE P_{in}(\lambda) \lambda d\lambda \quad (2.6)$$

where, $J_{sc}$ is the short-circuit current density, $h$ the Planck's constant (J.s), $c$ the velocity of light (m.s$^{-1}$), and $e$ the electronic charge in Coulomb.

A high value of *EQE* does not guarantee good photovoltaic energy conversion, but it is essential. The quantum conversion efficiency of the solar cells is usually much lower than 100% due to the losses associated with reflection of incident photons, their imperfect absorption by the photoactive material and recombination of the charge carriers before they reach the electrodes. Additionally, there are electrical resistance losses in the cell and in the external circuit due to both of electrodes and wire connections. *EQE* is also equal to the multiplication of all the efficiencies in the energy transfer processes [19]:

$$EQE = \eta_{abs} \eta_{diff} \eta_{tc} \eta_{tr} \eta_{cc} \quad (2.7)$$

Where, $\eta_{abs}$ is the photon absorption efficiency. The last four parameters, namely $\eta_{diff}$, $\eta_{tc}$, $\eta_{tr}$ and $\eta_{cc}$, are the internal quantum efficiency (*IQE*), which represents the efficiencies of the exciton diffusion process, the hole–electron separation process, the carrier transport process, and the charge collection process, respectively. The most effective way to improve the $J_{sc}$ is to enlarge $\eta_{abs}$. Therefore, OSCs materials need not only to absorb the photons at the maximum irradiance but also to have a broad absorption spectrum and high absorption coefficient.

## 5. Approaches to Improve Organic Solar Cells

It was previously illustrated that upon the absorption of light, excitons are formed in the organic photoactive layer, followed by the exciton diffusion and dissociation, which occurs at the donor/acceptor interface via an ultra-fast charge transfer between the LUMOs of donors and acceptors. Subsequently, the separated free electrons and holes transport through their individual percolating pathways, and then they are extracted by the corresponding electrodes. Therefore, the overall improvement in efficiency and the devices performance can be approached by enhancing various factors participating during the processes of light absorption, exciton diffusion/dissociation, charge transport, and charge collection.

One major obstacle in OSCs is how to make excitons dissociate effectively into free charge carriers. Another challenge, which is common in both of organic and inorganic solar cells, is the full collection of photogenerated charge carriers by the correspondence electrodes (negative and positive electrodes). In organic, especially disordered materials, carrier mobility is several orders of magnitude smaller than that in crystalline inorganic semiconductors. This imposes restrictions on the maximum thickness of organic photovoltaic devices and makes them to have very thin active layers (in nanometers scale). Furthermore, organic semiconductors suffer from imbalance of electron and hole mobilities within the same material. Accumulation of less mobile charge carriers in the bulk structure will hamper charge collection at the electrodes and thereby drastically reduce the solar cell efficiency. Another problem is difficulties in efficient harvest of excitons. Because of a large binding energy, intrinsic dissociation of excitons into free carriers is virtually impossible. Due to this, diffusion of excitons towards either charge transfer centers or donor/acceptor interfaces is a prerequisite for charge photogeneration [44]. Therefore, different practical approaches are needed to be undertaken to overcome the above mentioned bottlenecks in front of the OSCs. The next subsections will provide an overview of these approaches.

### 5.1 Bulk Heterojunction Structure

The idea behind bilayer heterojunction is to use two materials with different electron affinities (LUMO) and ionization potentials [45]. By this, favorable exciton dissociation is obtained; the electron will be accepted by the material with the higher electron affinity while the hole by the material with the lower ionization potential [8]. The main drawback of this concept resides in the rather short diffusion length of excitons (5–15 nm) [46]. Indeed, only those excitons that are created within a distance from the donor–acceptor (D-A) interface shorter than their diffusion length may contribute to the photocurrent generation [44]. This limits the photocurrent and hence the overall performance of bilayer organic solar cells. To overcome this limitation, the surface area of the D-A interface needs to be increased. This can be achieved by creating a mixture of donor and acceptor materials with a nanoscale phase separation resulting in a three-dimensional interpenetrating network; the "bulk heterojunction" [47] as shown in Figure 2.9. This suggests the goal of achieving a larger interfacial area between the electron- and hole-transporting materials. The photocurrent achieved by bulk heterojunction devices are up to several milliamperes per square centimeter, improving drastically the efficiencies of bilayer cells [32, 37]. Among these PV cells, solution-processed bulk heterojunction with donor–acceptor blends sandwiched between the anode and cathode are the most promising alternative to realize large-scale solar cell production [10, 20].

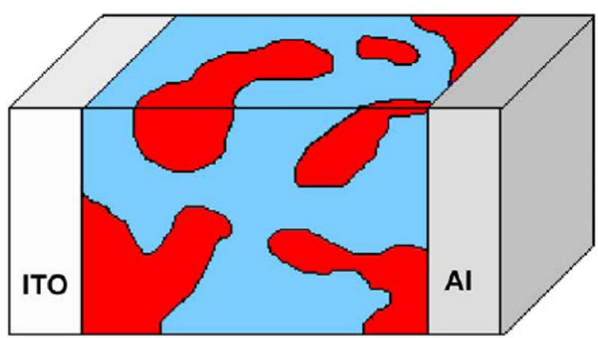

Figure 2.9: Bulk heterojunction structure between ITO and Al electrodes.

The bulk heterojunction device is similar to the bilayer device with respect to the D-A concept, but it exhibits a vastly increased interfacial area dispersed throughout the bulk. The bulk heterojunction requires percolated pathways for both phases throughout the volume, i.e., a bicontinuous and interpenetrating network. Therefore, the nanoscale morphology in the bulk donor/acceptor active layer is more complicated to be controlled well in terms of the phase separation between the donor and acceptor regions and hence possibly the leakage current persistence due to the insufficient contact of the acceptor phase with the cathode and donor phase with the anode electrodes.

**5.2 Multilayer and Tandem Structures**

As each organic material has a unique band gap, broad sunlight spectra cannot be harvested efficiently by a single layer. Therefore, multiple layers and stacked tandem cell structure, in which each layer absorbs a different light wavelength, can mostly resolve the limited absorption problem. Organic materials have higher extinction coefficient than that of inorganic materials. Therefore, about 300 nm film is thick enough to absorb the most incident light [19]. However, the thickness is ultimately limited by the short exciton diffusion length and low charge carrier mobility [47], while the balance between the light absorption and the charge transport plays a crucial role in the efficiency improvement. As a result, the optimized thickness for most of OSCs is less than 100 nm [19]. Several materials and device structures have been developed to obtain high short-circuit photocurrent density ($J_{sc}$), e.g., bulk heterojunction structure that was mentioned previously [9, 37], the concept of tri-layer organic p–i–n junction [48], in which i-layer is a co-deposited layer of two different organic semiconductors. Co-deposited layers have a vast number of heteromolecular (donor–acceptor) contacts acting as efficient photocarrier generation sites [46] (see Figure 2.11-(a)). Additionally, tandem structure as an effective approach to enhance the light harvesting by means of stacking multiple cells with complementary absorption spectra has also been proposed. The limits to power conversion efficiency and photovoltage can be breached through fabrication of tandem solar cells [49-51]. A tandem solar cell consists of two stacked solar cells made from materials with different optical gaps. Initially, light is absorbed by the higher-gap cell lower energy photons pass through the higher gap device and then photons are absorbed by the second cell. There is a conductive layer connecting the two cells accordingly, which works as a site for charge recombination. However, in terms of device fabrication, there are difficult tasks such as optimizing the layers thickness, selection of suitable materials, and recombination site that have to be treated with a great caution. Figure 2.11-(b) shows a representative tandem solar cell with two stacked cell structure having a gold metal (Au) layer as the conductive layer to provide the recombination site.

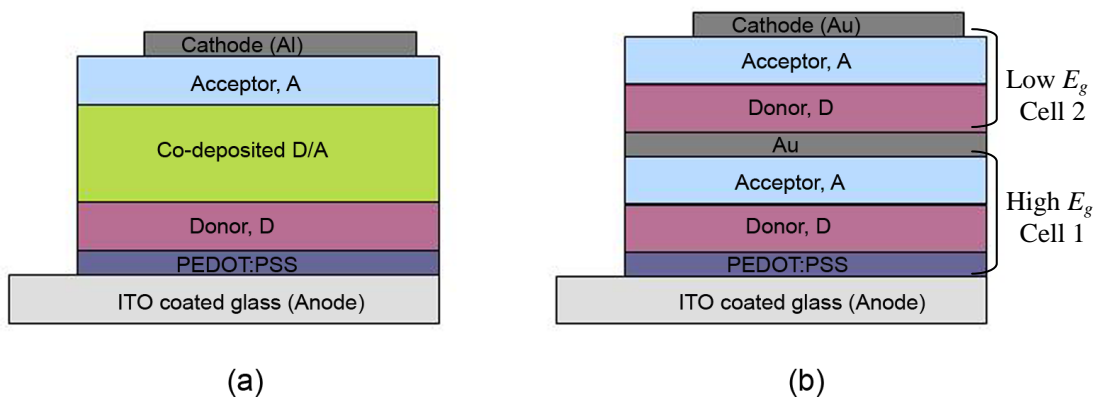

Figure 2.11: The structure of (a) multilayer organic p-i-n solar cells and (b) organic tandem solar cells.

**5.3 Exciton Blocking Layer**

As revealed in Figure 2.9, the active layer of bulk heterojunction structure is sandwiched between the anode and cathode electrodes, in which both donor and acceptor materials are in direct contact with the electrodes. Hence, it is possible for the acceptor material to transfer electrons to the hole-collecting anode and for the donor to transfer holes to the electron-collecting cathode, thereby resulting large leakage current and decreased PV performance of the cell. In order to tackle these problems, interfacial buffer layers are mostly inserted between the active layer and electrodes to enhance the collection of the photogenerated charges and to reduce the leakage current [10]. In a study in which an exciton blocking layer (EBL) of bathocuproine (BCP) was incorporated, photovoltaic properties of PV devices based on pentacene/PCBM were investigated [52]. It was seen that a thin layer of BCP has improved *EQE* and charge carrier collection of the devices. More recently, tris (8-hydroxyquinolinate) aluminium (Alq3) was also used between the cathode electrode and acceptor materials [53, 54] (see Figure 2.12). It was seen that this has led to increase in both efficiency and stability of the devices.

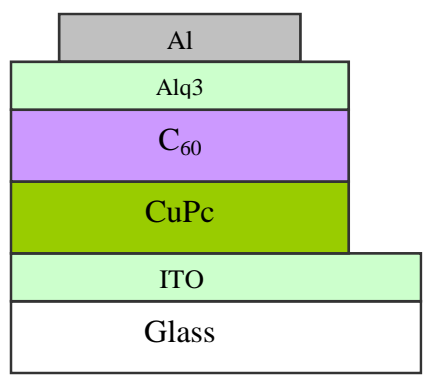

Figure 2.12: OSC device incorporating exciton blocking layer of Alq3 between the cathode electrode and acceptor material.

Similar to the cathode interface, atoms from the anode can react with the organic material. Indium atoms from indium tin oxide (ITO) anode were found to diffuse into the organic layer where it acts as trapping site for the charge carriers [8]. One strategy that is used to minimize indium and oxygen diffusion is to put an interfacial hole-transporting layer, such as poly(3,4-ethylenedioxythiophene):poly(4-styrenesulfonic) acid (PEDOT-PSS), between ITO and the active material. This layer also serves to smooth out the uneven surface of ITO and provides larger injection of holes into the anode electrode. As shown in Figure 2.13, the insertion of a thin PEDOT:PSS layer (≈35 nm) between the polymer blend and anode electrode, was seen to increase the overall efficiency of the devices significantly [37].

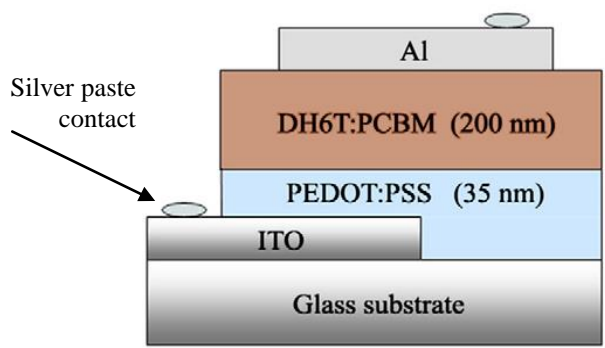

Figure 2.13: OSC device incorporating exciton blocking layer of PEDOT:PSS between the anode electrode and acceptor material.

**5.4 Double Cable Polymer**

It is clear that the control of morphology in dispersed bulk heterojunction devices is a critical point. The degree of phase separation and domain size depend on solvent choice, speed of evaporation, solubility, miscibility of the donor and acceptor, annealing temperature, etc. One strategy towards increasing the D-A phase control is to covalently link donor and acceptor to get some sort of polymers called double cable polymers. Recently, researchers [55-57] synthesized a pendant fullerene moieties and polythiophene backbone with covalently bound tetracyanoanthraquino-dimethane (TCAQ) moieties (donor–acceptor double-cable polymer), respectively, for utilization in organic solar cells. Even though the synthesized soluble cable polymers have shown some promising results, the fully optimized double-cable polymers are still not achieved. As the complexity of the designed systems increase, the more critical it becomes to optimize design parameters. The realization of effective double cable polymers will bring the D–A heterojunction at a molecular level. Figure 2.14-(a) shows a schematic representation of this system.

An alternative approach to double-cables polymer is block copolymers consisting of a donor and acceptor block as shown in Figure 2.14-(b). In general, block copolymers [58, 59] are well recognized for phase separation and ordered domains formation, similar to those of the double-cable polymers. [60] synthesized a block copolymer consisting of an electron acceptor block and an electron donor block through atom-transfer radical addition with the objective of enhancing the photovoltaic efficiency of the PPV-$C_{60}$ system (PPV = poly(*p*-phenylenevinylene)). Since the solubility of such complicated structures is very limited, the practical handling for device fabrication is cumbersome [8].

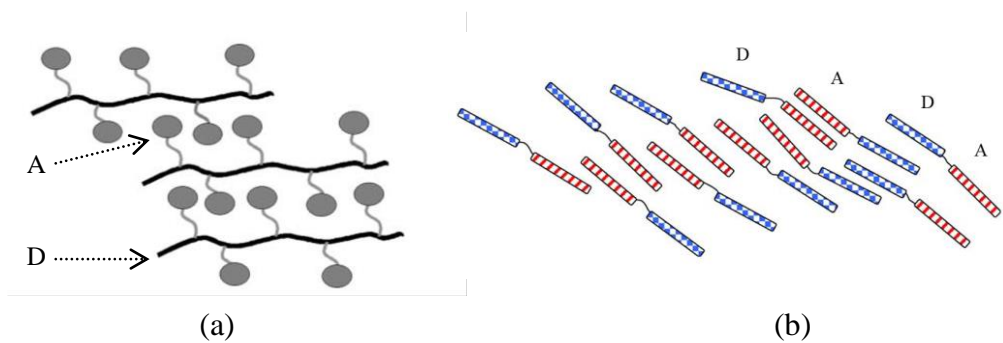

Figure 2.14: Shows (a) schematic representation of a realistic double-cable polymer where interchain interactions are considered, and (b) self-assembled layered structure of di-block copolymers.

**5.5 Energy Bands Alignment**

The discovery of semiconducting conjugated polymers stimulated the research field of organic electronics, thereby developing a variety of organic based devices such as, solar cells, light-emitting diodes, field-effect transistors, and memory devices. The emergence of the fields can be tracked back to the mid-1970 when Shirakawa reportedly prepared the first polymer (polyacetylene) by accident, subsequently Heeger and MacDiarmid discovered that the polymer would undergo an increase in conductivity of 12 orders [61]. The most important functionality of the organic materials is the large polarizability of their extended $\pi$-conjugated electron systems formed by the delocalization of the $p_z$-orbitals of the carbon atoms [62]. Due to this basic functionality, and upon the absorption of sun light, these materials can show the photo-induced charge carriers and transport properties by hopping process along their conjugated backbone. Therefore, most of the organic materials have attracted considerable attention to be exploited in the fabrication of electronic and optoelectronic devices [63]. In particular, two types of materials are usually selected as photovoltaic active layers in the fabrication of OSCs. The first layer must be conductive to holes. This is referred to as donor, while the second layer is conductive to electrons and known as acceptor. Regardless of that the device structure is a bilayer or bulk heterojunction, it is of great importance that the HOMO and LUMO energy levels of the donor-acceptor (D-A) system are matched well to facilitate efficient exciton generation, dissociation and charge transport conduction, as discussed earlier in Section 2.2. A basic description of the photoinduced charge transfer between a donor (D) molecule and an acceptor (A) molecule can be described as follows [64]:

Step 1: $D + A \rightarrow D^* + A$ (photoexcitation of D)

Step 2: $D^* + A \rightarrow (D-A)^*$ (excitation delocalized between D and A)

Step 3: $(D-A)^* \rightarrow (D^{\delta+} - A^{\delta-})^*$ (polarization of excitation: partial charge transfer)

Step 4: $(D^{\delta+} - A^{\delta-})^* \rightarrow (D^{\bullet+} - A^{\bullet-})$ (ion radical pair formation)

Step 5: $(D^{\bullet+} - A^{\bullet-}) \rightarrow D^{\bullet+} + A^{\bullet-}$ (complete charge separation)

Table 2.2 shows some representative electron donors and acceptors including their HOMO and LUMO energy levels with their molecular structures. Fullerenes are considered the best electron acceptors so far. This is because of: (i) ultrafast (~50 fs) photoinduced charge transfer that is happened between the donors and fullerenes; (ii) fullerenes exhibit high mobility, for example, $C_{60}$ has shown field effect electron mobility of up to 1 cm$^2$ V$^{-1}$ s$^{-1}$; (iii) fullerenes show a better phase segregation in the blend films [19, 64]. Among all the organic donor materials concerned, sexithiophene (6T) films show the highest mobility when they are used as hole transport layers [65, 66]. Upon improving the solubility of sexithiophenes by the addition of hexyl side chains to $\alpha$-sexithiophene (6T) [67] has provided us with $\alpha,\omega$-dihexyl-sexithiophene (DH6T) organic semiconductor which is characterized by reasonable field-effect mobility reaching as high as 0.1 cm$^2$/V.s [68, 69].

Table 2: The nomenclature, molecular energy levels, and structure of some representative organic donor and acceptor materials.

| | Nomenclature | HOMO (eV) | LUMO (eV) | Molecular structure |
|---|---|---|---|---|
| Donor | Poly [2-methoxy-5-(2'-ethyl-hexyloxy)-1,4-phenylene vinylene][a] | 5.2 | 2.8 | 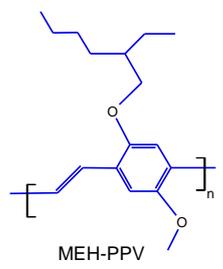 MEH-PPV |
| | Poly (3-hexylthiophene)[a] | 4.8 | 2.7 | |
| | | | | 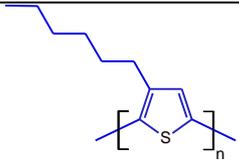P3HT |
| | Poly[2,6-(4,4-bis-(2-ethylhexyl)-4H-cyclopenta[2,1-b;3,4-b']dithiophene)-alt-4,7-(2,1,3-benzothiadiazole)][b] | 4.9 | 3.5 | 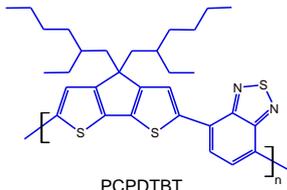PCPDTBT |
| *Donor* | α,ω-dihexyl-sexithiophene[c] | 5.2 | 2.9 | 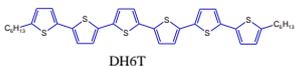DH6T |
| *Acceptor* | [6,6]-phenyl-$C_{61}$ butyric acid methyl ester (PCBM)[b] | 6.0 | 3.9 | 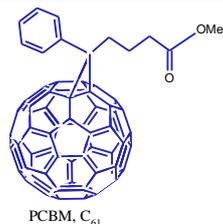PCBM, $C_{61}$ |
| | [6,6]-phenyl-$C_{70}$ butyric acid methyl ester (PCBM)[b] | 6.1 | 4.3 | 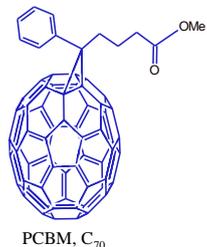PCBM, $C_{70}$ |
| | Tris (8-hydroxyquinolinate) aluminium[d] | 6.3 | 3.4 | 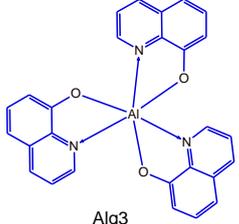Alq3 |
| | Tris (8-hydroxyquinolinate) gallium[d] | 5.8 | 3.0 | |

|  |  |  | 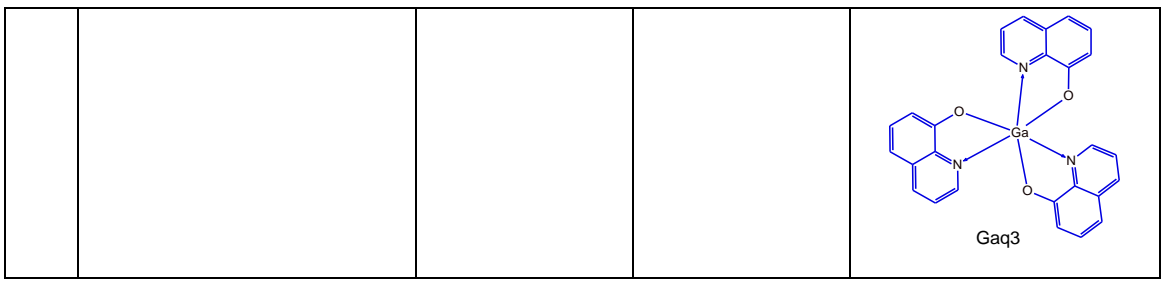 |
|---|---|---|---|
|  |  |  | Gaq3 |

(a) [38]
(b) [51]
(c) [70]
(d) [71]

Benefited by the high carrier mobility of DH6T and PCBM, and their appropriate energy band alignment for solar cells application [37], along with the possibility of tuning the optical band gap of DH6T by organometallic dopants [72], the arranged photoactive materials shown in Figure 2.15 can be used to formalize a viable DH6T/Mq3/PCBM ternary bulk heterostructure. A proper energy bands alignment of Gaq3 with the D-A components in fabricating solution-processed devices based on DH6T/Mq3/PCBM can be seen as shown in Figure 2.15. Nevertheless, Gaq3 was seen to show the most promising results compared to those of Alq3 [71, 73, 74], which will be discussed later in Chapter 3. Figure 2.15 shows the molecular energy levels of the DH6T/Gaq3/PCBM components (D-A-A), in which a satisfactory HOMO and LUMO energy bands alignment can be seen as one of the prerequisites in the OSCs fabrication.

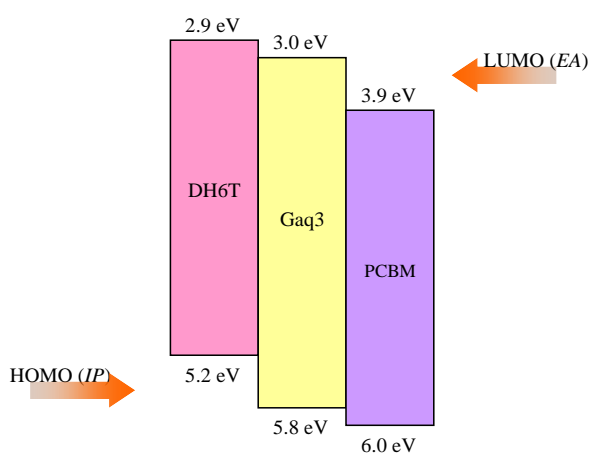

Figure 2.15: The correct HOMO and LUMO energy band alignment of DH6T, Gaq3 and PCBM, from left to right, respectively for the OSCs application.

## 6. Outline

The simple and low-cost fabrication process of organic solar cells makes them attractive candidates for generation of electricity. The performance enhancement of these devices is of prior request by the researchers in the field. This can be accomplished upon rigorous research studies performed through materials analysis to the devices fabrication and characterization, thereby realizing valuable strategies to improve the performance of these devices. The architecture of the active layer plays a vital role in defining the overall efficiency of OSCs. In this way, various approaches can be utilized to modulate the device components including heterojunction and multilayer structures, insertion of excitons blocking layers, designing the double cable or di-block copolymers along with the optimum selection for the HOMO and LUMO energy levels between the cells components.